\documentclass{appolb}

\usepackage{epsfig}
\begin{document}
\title{Probing deconfinement in the Polyakov-loop extended
Nambu-Jona-Lasinio model at imaginary chemical potential%
\thanks{Talk presented at ``Three Days on Quarkyonic Island'', HIC for FAIR
workshop and XXVIII Max Born Symposium, Wroc\l aw, 19-21 May 2011.}%
\thanks{YITP-11-95}
}
\author{Kenji Morita$^1$, Vladimir Skokov$^{2,3}$, Bengt Friman$^3$, and
Krzysztof Redlich$^4$
\address{$^1$Yukawa Institute for Theoretical Physics, Kyoto University,
Kyoto 606-8502, Japan, \\
$^2$Physics Department, Brookhaven National Laboratory, Upton, NY 11973, USA
$^3$GSI Helmholtzzentrum f\"{u}r Schwerionenforschung, D-64291 Darmstadt, Germany,\\
$^4$Institute of Theoretical Physics, University of Wroclaw, PL-50204 Wroc\l aw, Poland}
}
\maketitle
\begin{abstract}
 The phase structure of Polyakov-loop extended Nambu-Jona-Lasinio (PNJL)
 model is explored at imaginary chemical potential, with particular
 emphasis on the deconfinement transition. We point out that the
 statistical confinement nature of the model naturally leads to
 characteristic dependence of  the chiral condensate 
$\langle \bar{q}q  \rangle$ on $\theta=\mu_I/T$. We introduce a dual
 parameter for the deconfinement transition by making use of this
 dependence. By changing a four-fermion coupling
 constant, we tune the location of the critical endpoint of the deconfinement
 transition.
\end{abstract}
  
\section{Introduction}
Phase transitions in QCD have been extensively studied in lattice
quantum chromodynamics. While recent development enables us to
perform numerical simulations at physical quark masses, which revealed a
crossover nature of the QCD phase transition at finite temperature \cite{aoki06}, 
analyses at nonzero quark chemical potential $\mu$ have been limited to
small $\mu$ region due to the complex fermion determinant, 
known as the ``sign problem'' \cite{muroya03}. One of several methods circumventing this
problem is to use an imaginary chemical potential $\mu=i\mu_I$. 
Indeed, this method has provided transition lines in the  $T-\mu$ plane via
an analytic continuation from those obtained at imaginary $\mu$
\cite{forcrand02:_qcd,d'elia-lombardo,d'elia07}. Moreover, it has been known that there is a
phase transition specific to the imaginary chemical potential
characterizing the deconfinement phase at high temperature
\cite{roberge86:_gauge_qcd}.  Rich phase structures later found in the
lattice simulations provide a testing ground for understanding the
nature of phase transitions in QCD
\cite{forcrand:_const_qcd,bonati:_rober_weiss_endpoin_in_n_f_qcd,Chen, Nagata}.
Those properties give constraints on model studies which can be extended
to real $\mu$. In this work, we study the phase structure of the
Polyakov-loop extended Nambu-Jona-Lasinio (PNJL) model
\cite{fukushima04:_chiral_polyak,ratti06:_phases_qcd} which satisfies
fundamental symmetries of QCD relevant for phase transitions at
imaginary chemical potential. Focusing on the deconfinement transition,
we show that the ``statistical confinement'' feature of the model
naturally leads to characteristic behaviors of the order parameters while details
depend on the choice of the Polyakov loop potential. We discuss dual
parameters to characterize the phase transitions. Finally, we point out the
existence of the critical endpoint (CEP) associated with the
deconfinement transition at imaginary chemical potential and clarify the
relation between its location and the chiral phase transition.

In the next section, we will give a brief introduction of the model. We
will discuss the characteristic behavior of the order parameters as well
as the dual parameters in Sec.~\ref{sec:orderparameter}. The critical
endpoint of the deconfinement transition will be discussed in
Sec.~\ref{sec:cep} and Section \ref{sec:summary} is devoted to the
summary. More details can be found in Ref.~\cite{morita11:_probin_decon_in_chiral_effec}.

\section{PNJL model at imaginary chemical potential}

The Lagrangian of the two-flavor PNJL model is given by 
\begin{equation}
 \mathcal{L} = \bar{q}(i \gamma_\mu D^\mu -m_0 )q +
  G_s[(\bar{q}q)^2+(\bar{q}i\gamma_5 \vec{\tau}q)^2]-\mathcal{U}(\Phi[A],\Phi^*[A];T).
\end{equation}
The model is an extension of the NJL model, which is an
effective model of chiral properties of QCD
\cite{nambu61:_NJLI,hatsuda94:_qcd_lagran},  such that
quarks couple with background gluonic fields described by a $Z(3)$
symmetric effective potential $\mathcal{U}$ which takes care of
confinement. 
In the covariant derivative 
$D^\mu = \partial^\mu - i A^\mu$, only the temporal components of 
$A_0= gA_0^a \lambda^a/2$ is included. The effective potential
$\mathcal{U}$ is expressed in terms of the traced Polyakov loop and its conjugate,
$\Phi = \langle {\rm Tr}_c L \rangle/3$ and
 $\Phi^* = \langle {\rm Tr}_c L^\dagger \rangle/3$, respectively. 
This coupling between quarks and gluons leads to an almost simultaneous
crossover of the chiral and deconfinement transitions at finite
temperature, of which order parameters are chiral condensate
$\sigma\equiv\langle\bar{q}q\rangle$ and the Polyakov loop $\Phi$
\cite{fukushima04:_chiral_polyak}, provided the Polyakov loop potential
$\mathcal{U}$ yields a first order transition at $T_0=270$  MeV in
accordance with pure $SU(3)$ lattice calculations.
Two functional forms of $\mathcal{U}$, which reproduce the thermodynamic quantities
obtained in pure $SU(3)$ lattice gauge theory \cite{Boyd}, have been used. 
One has a polynomial form
\begin{equation}
 \frac{\mathcal{U}_{\rm poly}}{T^4} = -\frac{b_2(T)}{2}\Phi^* \Phi
  -\frac{b_3}{6}[\Phi^3+(\Phi^*)^3]+\frac{b_4}{4}(\Phi^* \Phi)^4\label{eq:upol}
\end{equation}
with a set of parameters given in \cite{ratti06:_phases_qcd}.
The other is a logarithmic one \cite{roessner07:_polyak}
\begin{equation}
 \frac{\mathcal{U}_{\rm log}}{T^4} = -\frac{a(T)}{2}\Phi^* \Phi +
  b(T)\log\{1-6\Phi^* \Phi+4[\Phi^3 + (\Phi^*)^3]-3(\Phi^* \Phi)^2\}.\label{eq:ulog}
\end{equation}
The logarithm restricts possible values of $\Phi$ and $\Phi^*$ to the
so-called target space, since the argument of the logarithm must be
positive.

At imaginary $\mu$, the two Polyakov loop variables $\Phi$
and $\Phi^*$ are complex conjugate
\cite{sakai08:_polyak_nambu_jona_lasin}. 
Moreover, the partition function of the PNJL model at imaginary chemical
potential has been shown \cite{sakai08:_polyak_nambu_jona_lasin} to
have the same periodicity in $\theta=\mu_I/T$ as that of QCD, $Z(\theta+2\pi/3)=Z(\theta)$,
which was pointed out by Roberge and Weiss \cite{roberge86:_gauge_qcd} as a
remnant of $Z(3)$ symmetry.
Therefore we may express them by using a modulus and a phase
$\Phi = |\Phi|e^{i\phi}$ and $\Phi^* = |\Phi|e^{-i\phi}$. 

The thermodynamic potential in the mean field approximation reads
\begin{eqnarray} 
\Omega(T,V,\theta) &=& (G_s \sigma^2 + \mathcal{U})V -4V \int\frac{d^3 p}{(2\pi)^3} 
 \left[ 3(E_p-E_p^0) \right.\nonumber \\
+T\ln[\!\!\!&1&\!\!\!+3|\Phi|e^{i(\theta+\phi)-\beta E_p} +3 |\Phi|
e^{i(2\theta-\phi)-2\beta E_p}+e^{3i\theta-3\beta E_p}] \nonumber \\
+T\ln[\!\!\!&1&\!\!\!+3|\Phi|e^{-i(\theta+\phi)-\beta E_p} +3 |\Phi|
e^{i(\phi-2\theta)-2\beta E_p}+e^{-3i\theta-3\beta E_p}]]\label{eq:omega}
\end{eqnarray}
where $E_p=\sqrt{p^2+M^2}$, $E_p^0=\sqrt{p^2+m_0^2}$, and
$M=m_0-2G_s\sigma$. The first term in the momentum integral is a
divergent vacuum term, which is regularized by a three-momentum cutoff
$\Lambda$. The cufoff and coupling are fixed to $G_s=5.498$ GeV$^{-2}$
and $\Lambda=0.6315$ GeV so as to reproduce the vacuum pion mass and
pion decay constant with $m_0=5.5$ MeV. In the following we mainly focus
on the result in the chiral limit $m_0=0$ to preserve the chiral symmetry in
the Lagrangian.
The chiral condensate $\sigma$ serves as an order
parameter for the chiral phase transition. The order
parameters are determined by the minimum of the potential which is
obtained by solving the gap equation $\partial \Omega/\partial X_i=0$ with 
$X_i = M, |\Phi|, \phi$.

\section{Behavior of the order parameters}
\label{sec:orderparameter}
\subsection{Order parameters at imaginary chemical potential}

\begin{figure}[!t]
 \epsfig{file=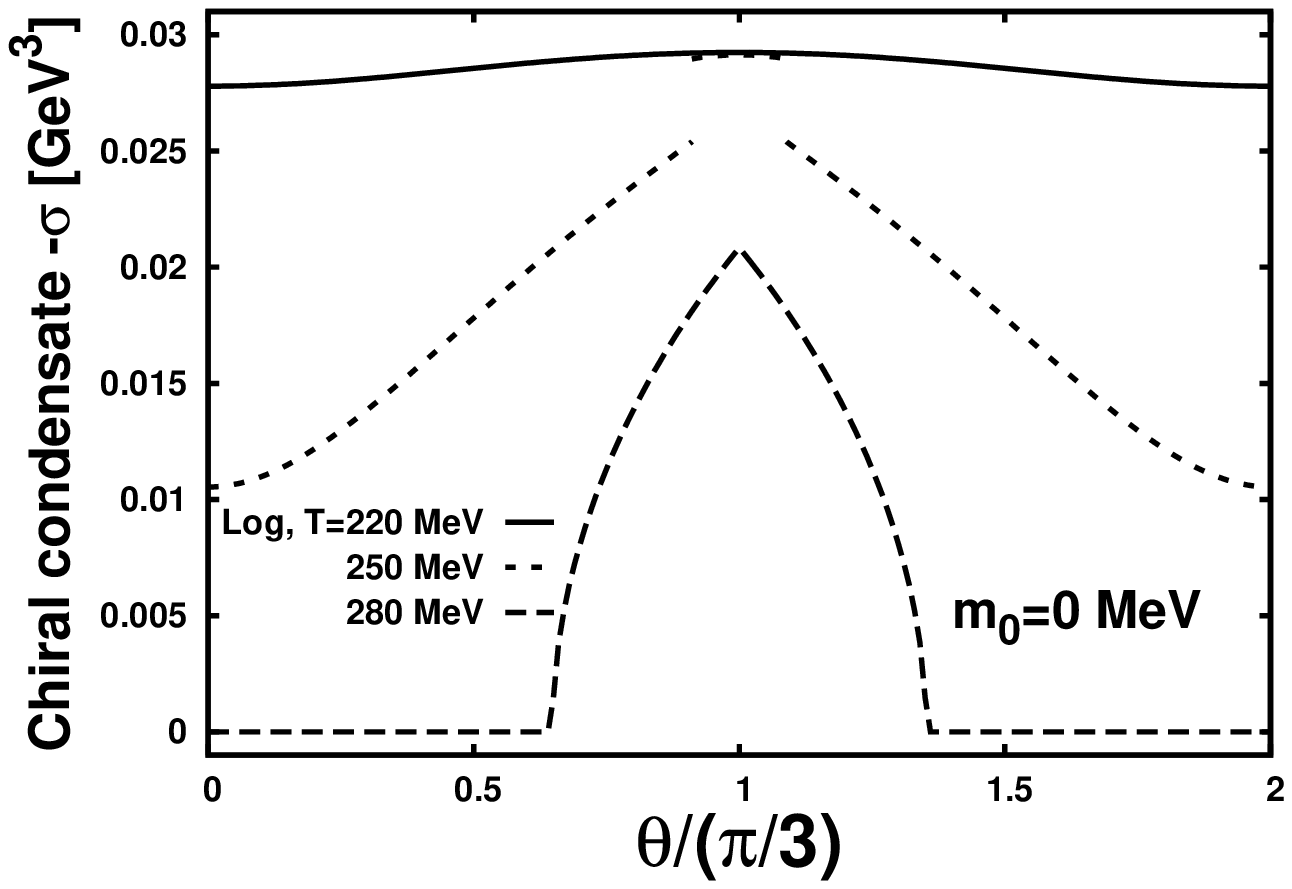,width=0.45\textwidth}
 \epsfig{file=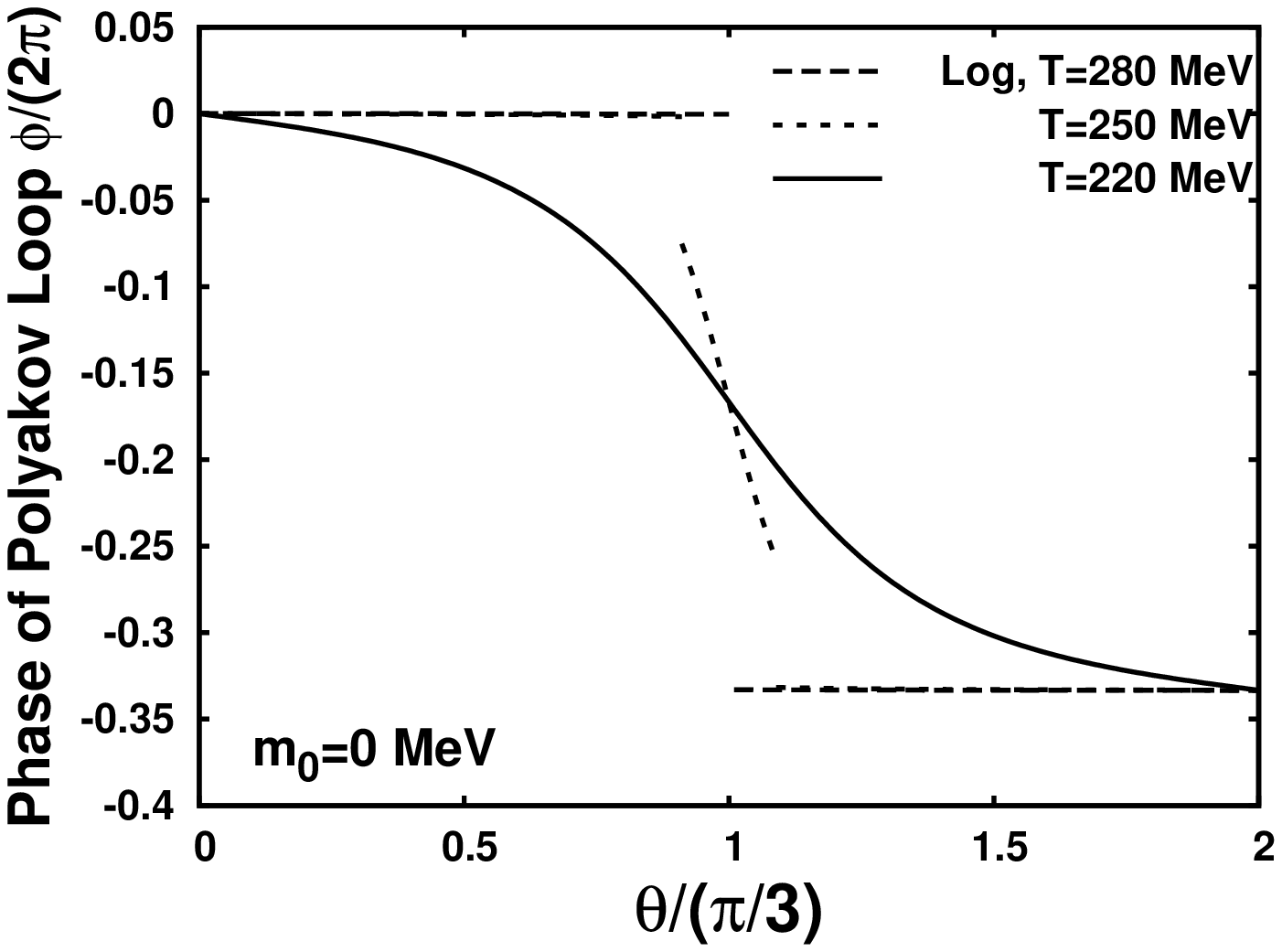,width=0.45\textwidth}
 \caption{Left : Chiral condensate for various temperatures as functions
 of $\theta$. Right : Phase of the Polyakov loop $\phi$. Both results
 are in the chiral limit $m_0=0$ and for the logarithmic potential.}
 \label{fig:sigma}
 \epsfig{file=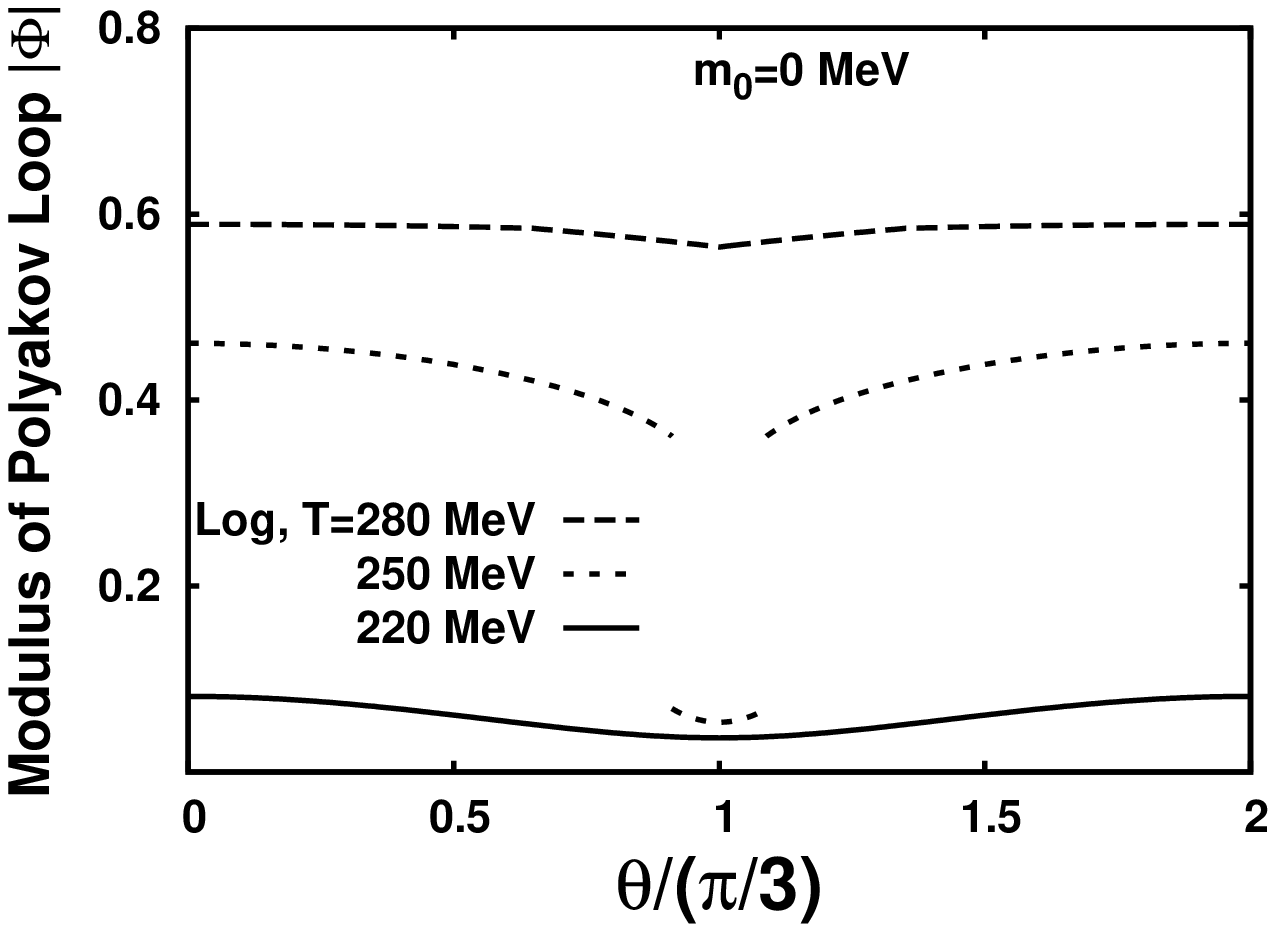,width=0.45\textwidth}
 \epsfig{file=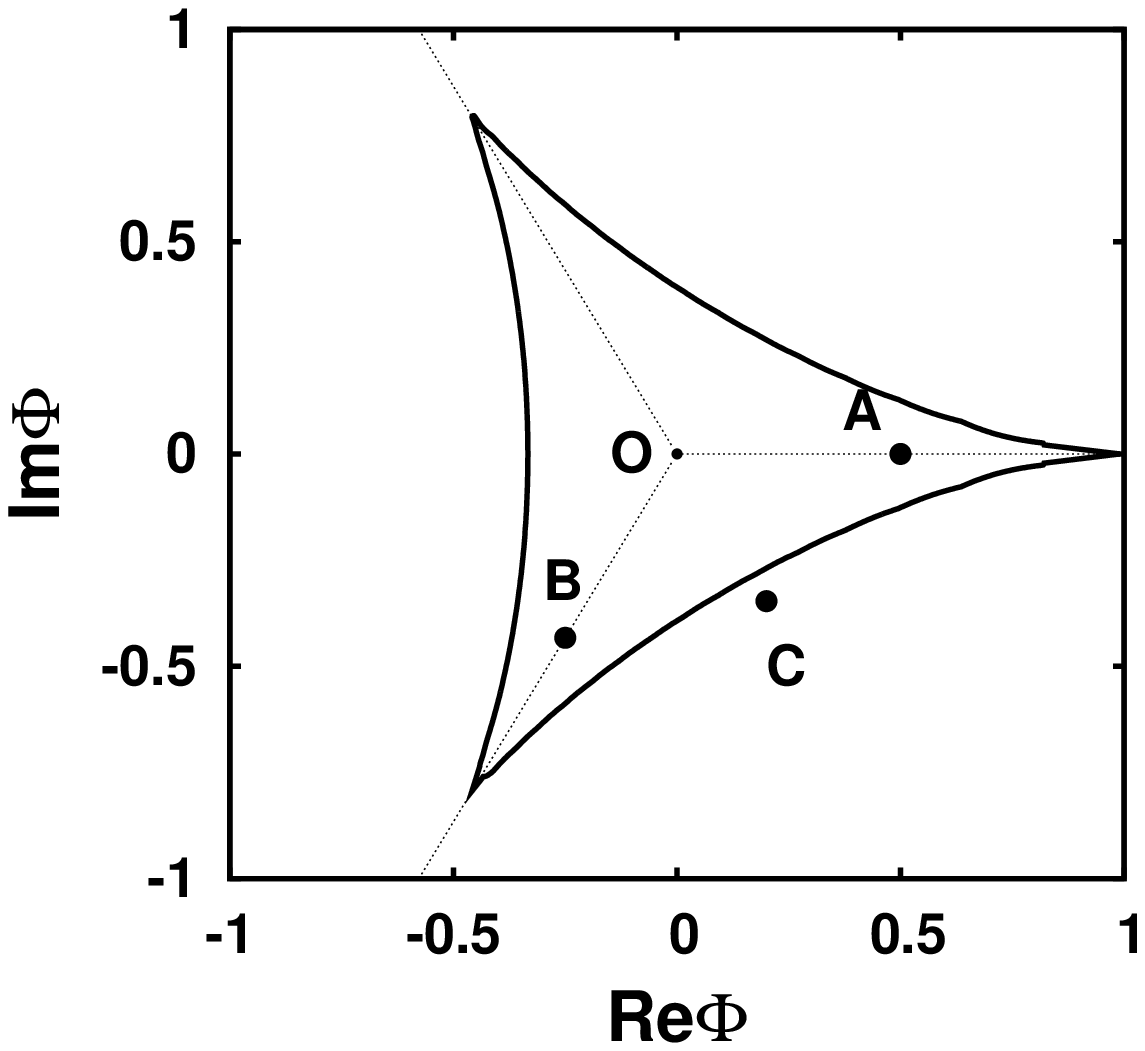,width=0.4\textwidth}
 \caption{Left : Modulus of the Polyakov loop $|\Phi|$. Right : target
 space of the Polyakov loop on complex $\Phi$ plane. A region inside the
 solid lines denote the target space of the logarithmic potential in
 which the argument of the logarithm is positive.}
 \label{fig:moduls_target}
\end{figure}

First we consider two extreme limits in order to see characteristic
$\theta$ dependences of $\sigma$ which has the same periodicity $2\pi/3$
as $\Omega$.
Expanding Eq.~(\ref{eq:omega}) for small $e^{-\beta E_p}$, we have a gap
equation at small $|\Phi|$ limit in which only a term proportional to
$\cos3\theta$ remains with a small magnitude $\sim e^{-3 \beta E_p}$
indicating the statistical confinement. 
This dependence naturally leads to the periodicity
$2\pi/3$ when $|\Phi|$ is negligible and chiral symmetry is broken. 
On the other hand, when $|\Phi|\simeq 1$, the model reduces to the NJL
model except for the coupling of $\phi$ with $\theta$, as seen in
Eq.~(\ref{eq:omega}). In this case, the apparent $\theta$ dependence is governed by
$\cos\theta$ as a consequence of deconfinement. Although this factor
does not match with the required periodicity $2\pi/3$, it is preserved
by a change of $\phi$, namely, the Roberge-Weiss transition.

We show the chiral condensate in the left panel of
Fig.~\ref{fig:sigma} obtained by numerically solving the gap equations.
One sees that $\sigma$ at low temperature ($T=220$ MeV) exhibits small and smooth
variation as a function of $\theta$, as discussed above. On the other
hand, one sees a cusp at $\theta=\pi/3$ and $T=280$ MeV.
This is a consequence of a Roberge-Weiss transition depicted in
the right of Fig.~\ref{fig:sigma}, in which the phase $\phi$ changes from 0 to
$-2\pi/3$, smoothly at low $T$ but discontinuously at high $T$. 
As a result, the required periodicity of $\sigma$ is preserved.
One also sees a second order chiral phase transition for $T=280$ MeV
in which the chiral symmetry is broken around $\theta=\pi/3$. This implies the
chiral critical temperature at imaginary $\mu$ is higher
than that of zero and real $\mu$. This can be also
understood from the gap equation for $\sigma$, since $\cos n\theta$ is
replaced by $\cosh n \beta \mu$ for real $\mu$.

While the above properties are independnent of the choice of
$\mathcal{U}$, there are some potential dependent features as follows.
In Fig.~\ref{fig:sigma} and the left of Fig.~\ref{fig:moduls_target},
one sees a discontinuity in the order parameters at the same $\theta$. 
This shows a first order deconfinement transition which exists only in
the case of the logarithmic potential (\ref{eq:ulog}). 
The polynomial potential (\ref{eq:upol}) exhibits smoother change
near phase transition. 
In the right of Fig.~\ref{fig:moduls_target}, the target space of the
Polyakov loop is displayed. Owing to the $Z(3)$ symmetry,
$\mathcal{U}$ has three degenerate minima at $T > T_0$. Putting quarks
into the system makes one of those minima favored. While ${\rm Im}\Phi=0$
is always chosen at $\theta=0$, ${\rm Im}\Phi \neq 0$ is favored at
imaginary chemical potential due to the coupling of $\theta$ and $\phi$
seen in Eq.~(\ref{eq:omega}). At low temperature where minimum of
$\mathcal{U}$ is close to the origin, the minimum of the effective potential
smoothly moves from $\phi=0$ to $\phi=-2\pi/3$ across $\theta=\pi/3$. 
At high temperature, however,
there is a potential wall which makes the transition from point A to
point B discontinuous. Since the polynomial potential does not have any
restriction of the target space, the minimum passes outside (C) the target space
near the RW transition.

\begin{figure}[t]
 \epsfig{file=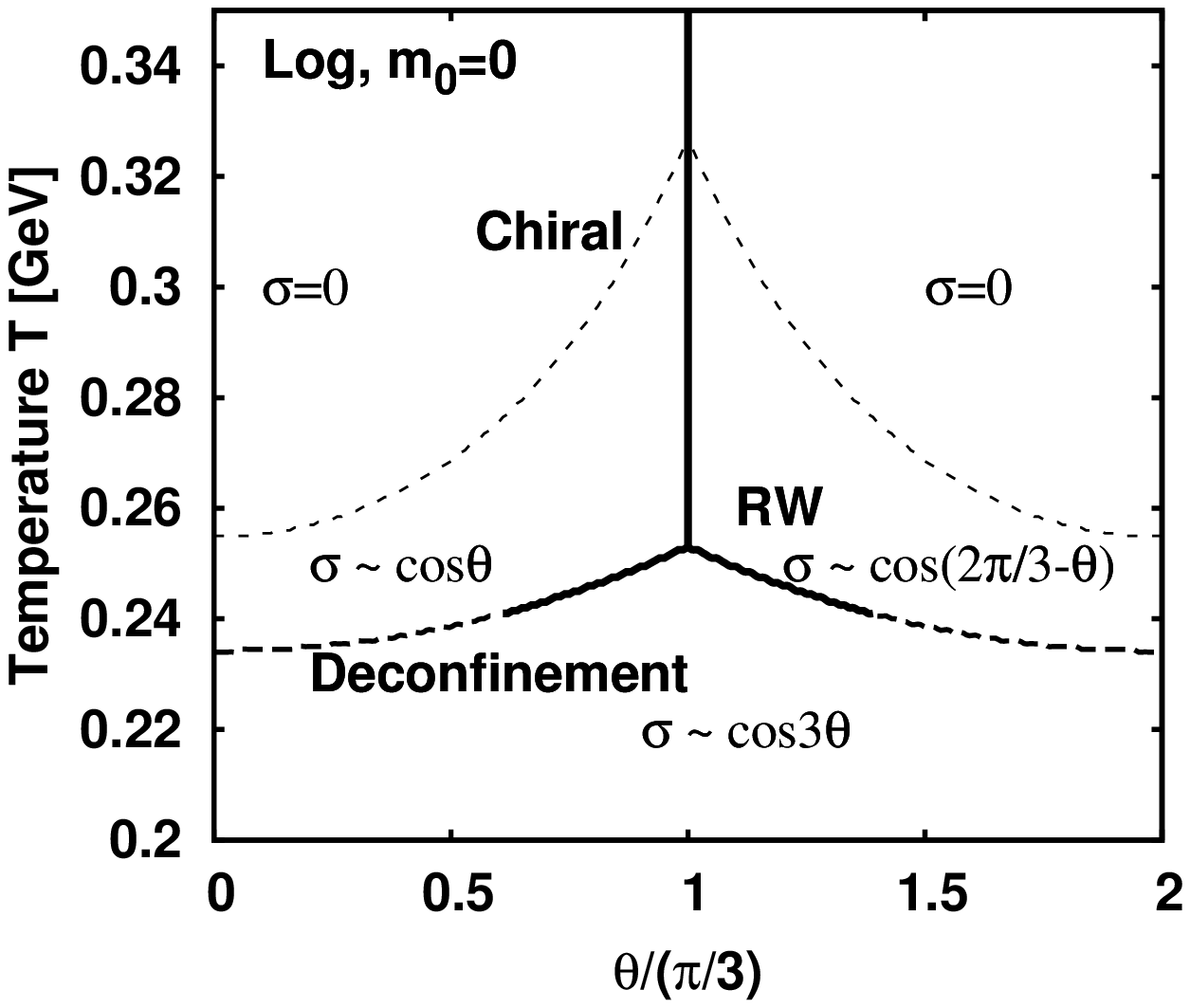,width=0.45\textwidth}
 \epsfig{file=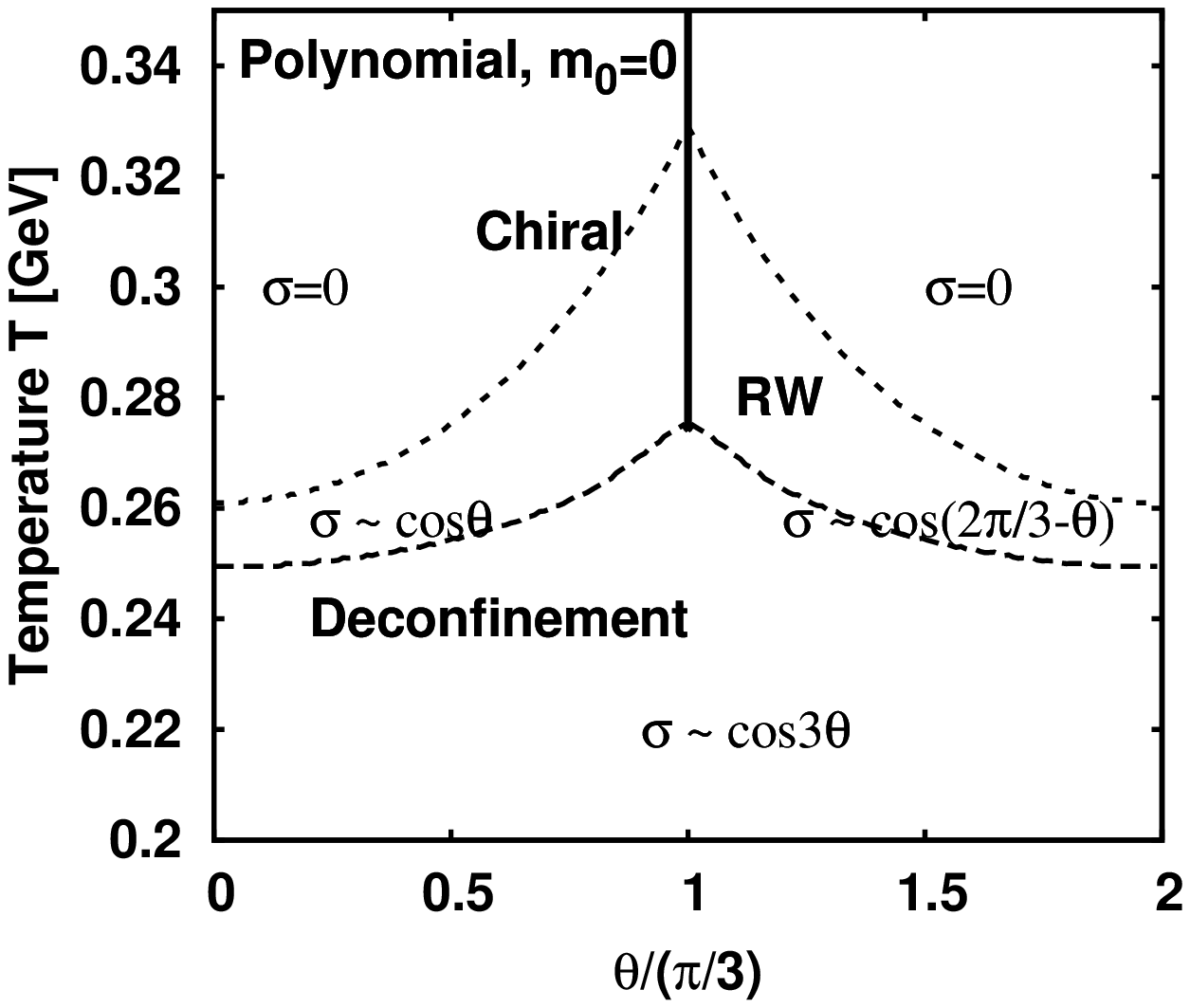,width=0.45\textwidth}
 \caption{Phase diagram on $T-\theta$ plane. Solid lines, dotted lines,
 and dashed lines stand for first, second, and crossover transitions,
 respectively.}
 \label{fig:pd}
\end{figure}
The phase diagrams shown in Fig.~\ref{fig:pd} summarize the behavior of the
 order parameters. One sees a first order deconfinement transition
and an associated critical endpoint (CEP) only for the logarithmic
potential. This also implies that the RW endpoint, where the first order RW
transition terminates, is a triple point. On the other hand, one sees a
 second order RW endpoint for
the polynomial potential. The properties of the RW endpoint in QCD might reflect
the nature of the QCD phase transition at real $\mu$
We refer to Refs.~\cite{bonati:_rober_weiss_endpoin_in_n_f_qcd}
and \cite{forcrand:_const_qcd} for recent calculations of $N_f=2$ and $N_f=3$ latttice QCD,
respectively. Especially it should be noted that the order of the RW
endpoint has a non-trivial bare quark mass dependence which cannot be
reproduced by chiral effective models. (See Sec.~\ref{sec:cep}) 
An improved model was proposed in Ref.~\cite{sakai:_10063408} to
reproduce this property. 

\subsection{Dual parameters for deconfinement}

\begin{figure}[t]
 \epsfig{file=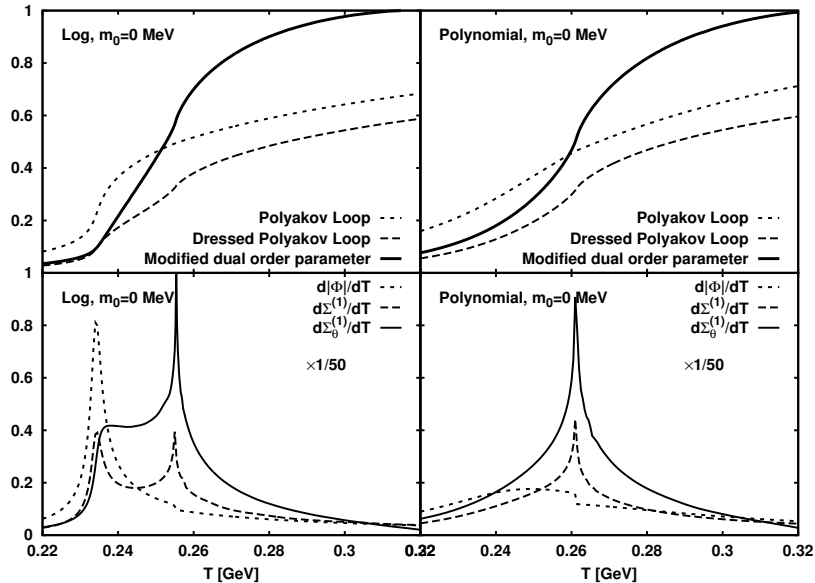,width=0.9\textwidth}
 \caption{Dual parameters compared with Polyakov loop. Top panels show
 $\Phi$, $\Sigma^{(1)}$ and $\Sigma_\theta^{(1)}$ for 
 $\mathcal{U}_{\rm  log}$ (left) and $\mathcal{U}_{\rm poly}$ (right)
 while bottom ones displays their derivatives with respect to temperature.}
 \label{fig:dual}
\end{figure}

It has been shown that information on the deconfinement is
encoded in $\theta$ dependence of $\sigma$. We can consider dual
parameters which characterize the deconfinement transition.
A dual parameter was introduced in \cite{bilgici08:_dual_polyak}.
By considering a twisted boundary condition for quarks 
$q({\mathbf{x}},\beta)=e^{i\varphi}q(\mathbf{x},0)$, one may define the
corresponding chiral condensate $\sigma(\varphi)$. Then the dual chiral
condensate $\Sigma^{(n)}$ reads
\begin{equation}
 \Sigma^{(n)}(T) =
  -\int_{0}^{2\pi}\frac{d\varphi}{2\pi}e^{-in\varphi}\left[
						      -\frac{1}{V}\left\langle{\rm
								   Tr}[(m_0+D_\varphi)^{-1}]\right\rangle
						     \right]
\end{equation}
While the twisted boundary condition is similar to introducing imaginary
chemical potential \cite{weiss87:_how}, it does not apply to the
background gauge field. Therefore, $\sigma(\varphi)$ has a periodicity
$2\pi$ and was calculated in a PNJL model by fixing the Polyakov loop at
its $\theta=0$ value \cite{kashiwa09:_dual_polyak_nambu_jona_lasin}. 
Particularly $\Sigma^{(1)}$ is called dressed Polyakov loop, since it
has the same transformation properties under $Z(3)$ and thus is expected
to serve as an order parameter of  the deconfinement transition.
Analogously, we consider a modified dual parameter which utilizes the characteristic
property of $\sigma(\theta)$,
\begin{equation}
 \Sigma_\theta^{(n)}(T) = \frac{3}{2\pi}\int_{-\pi/3}^{\pi/3}d\theta e^{-in\theta}\sigma(T,\theta).
\end{equation}
where we take the integration range $[-\pi/3,\pi/3]$, owing to the
periodicity of $\sigma(\theta)$.

We compare those dual parameters for $n=1$ with the Polyakov loop in
Fig.~\ref{fig:dual}, as well as their derivatives with respect to
temperature, of which peaks can be regarded as (pseudo)critical
temperatures. One sees that while dual parameters show a rapid increase
as seen in the Polyakov loop (top),\footnote{Dual parameters are normalized to
0 as $T\rightarrow 0$ and 1 as $T\rightarrow\infty$
\cite{morita11:_probin_decon_in_chiral_effec}.} their derivatives
exhibit different peak structures. The derivatives of the dual parameters have a peak at
the chiral transition temperature, independent of $\mathcal{U}$.  
As for the deconfinement, however, existence of the peak depends on
$\mathcal{U}$.
The dressed Polyakov loop exhibits a peak for the 
$\mathcal{U}_{\rm log}$ for which $|\Phi|$ shows stronger crossover than
$\mathcal{U}_{\rm poly}$. Moreover, the modified dual parameter exhibits
only a shoulder even for $\mathcal{U}_{\rm log}$. This result indicates
different sensitivity of the dual parameters to the chiral and
deconfinement transition. 

\section{Critical endpoint of deconfinement}
\label{sec:cep}

\begin{figure}[!t]
 \epsfig{file=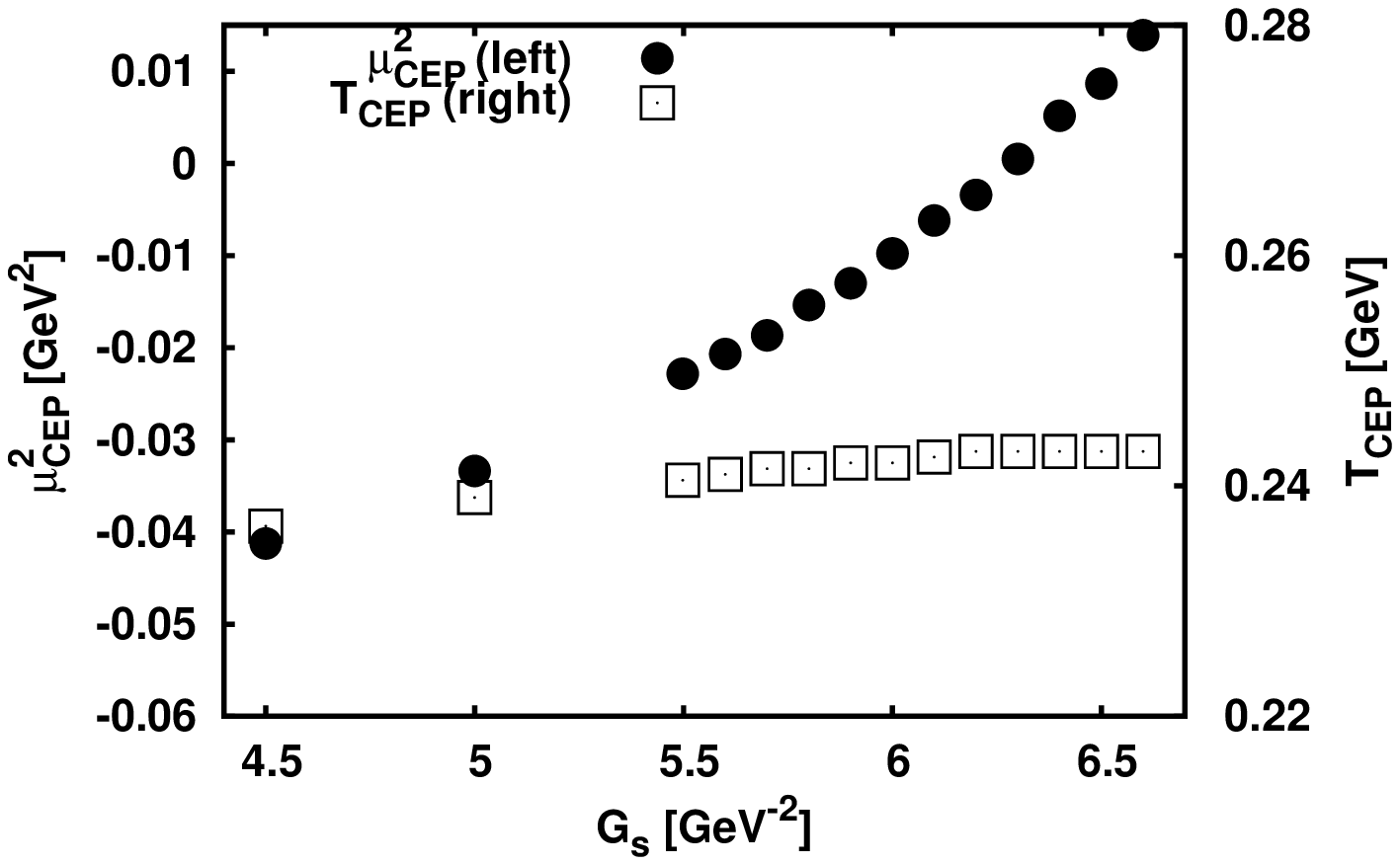,width=0.45\textwidth}
 \epsfig{file=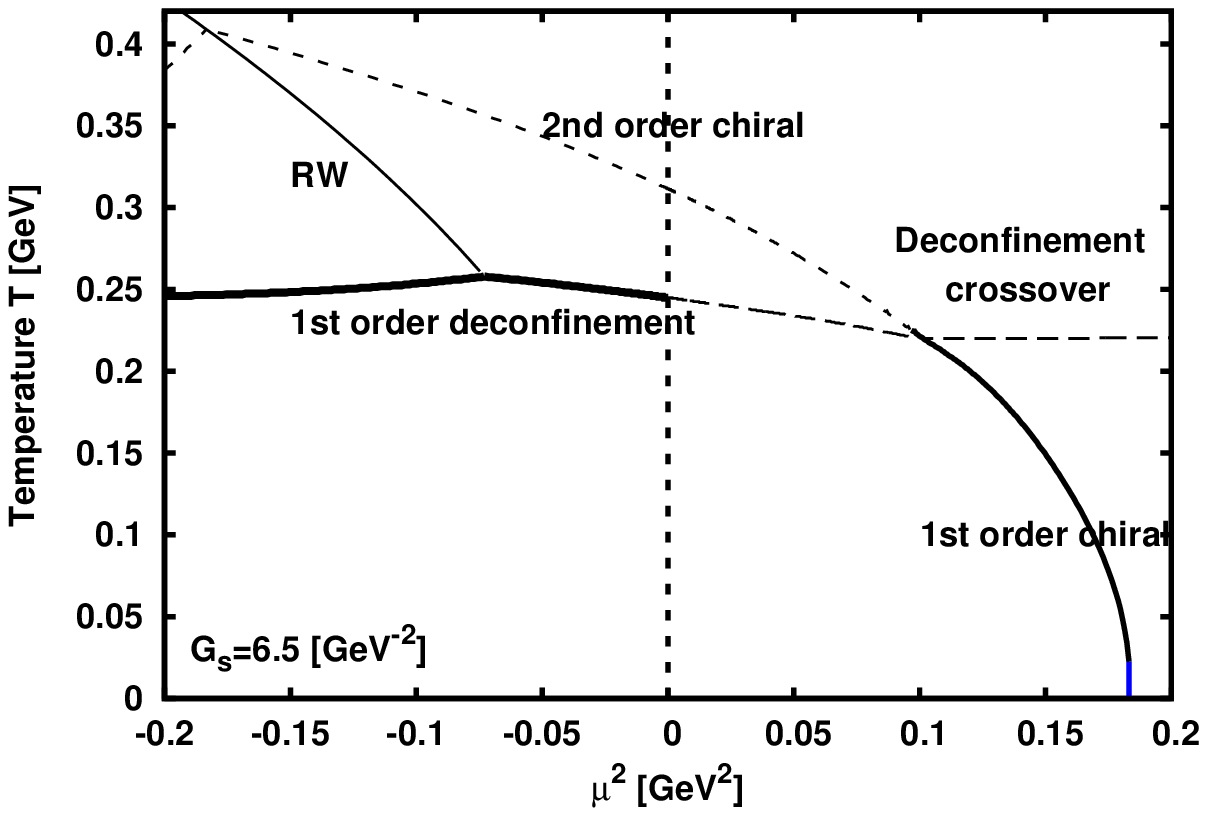,width=0.45\textwidth}
 \caption{Left : location of the deconfinement CEP 
 $(T_{\rm CEP}, \mu^2_{\rm CEP})$ as a function of $G_s$. Right : Phase
 diagram for $G_s=6.5$ GeV$^{-2}$ which gives $\mu_{\rm CEP}^2 > 0$.}
 \label{fig:deccep}
\end{figure}

Now let us turn to the deconfinement CEP found in the case of
$\mathcal{U}_{\rm log}$. Here we vary the four-fermion coupling constant
$G_s$ to preserve the chiral symmetry in the Lagrangian. Locations of the CEP are shown in
the left of Fig.~\ref{fig:deccep} for various values of $G_s$. One sees the squared critical chemical
potential $\mu_{\rm CEP}^2$ increases with $G_s$ to reach 
$\mu_{\rm CEP}^2=0$ around $G_s\simeq 6.3$ GeV$^{-2}$. 
In the right of Fig.~\ref{fig:deccep}, we also depict a phase diagram
for $G_s=6.5$ GeV$^{-2}$ in which the CEP exists at real chemical
potential. One sees that the first order deconfinement transition starting
from the RW endpoint (see Fig.~\ref{fig:pd}) is prolonged, while the
chiral critical line moves upward. The relation between these two
changes can be understood as follows. Since the Polyakov loop potential
$\mathcal{U}_{\rm log}$ has a first order phase transition at $T=T_0$, the model
results in the same transition when the effects of quarks are negligible
in thermodynamics. The contribution of quarks to thermodynamic potential
is essentially determined by the dynamical quark mass $M=m_0-2G_s\sigma$, not by the
current quark mass $m_0$, as seen in Eq.~(\ref{eq:omega}). When
dynamical quark mass becomes lighter around $T=T_0$, the deconfinement transition is
modified to a crossover one. 
As $G_s$ increases, the stronger coupling leads to a larger condensate
$|\sigma(T=0)|$ thus the dynamical quark mass becomes heavier. This appears as the modified chiral critical line in
the phase diagram at $G_s=6.5$ GeV$^{-2}$ and the resultant dynamical
quark mass is heavy enough to recover the first order the deconfinement
transition. At the reference value of $G_s$, the imaginary chemical
potential
weakens the thermal terms by $\cos n\theta$ in the thermodynamic potential thus resembling a
heavier quark mass which yields the CEP and a first order transition. 
While the above consideration is completely independent of the form of
$\mathcal{U}$, quantitative features such as the value of dynamical quark
mass which makes the transition first order depend on the choice of $\mathcal{U}$.

\begin{figure}[!t]
 \epsfig{file=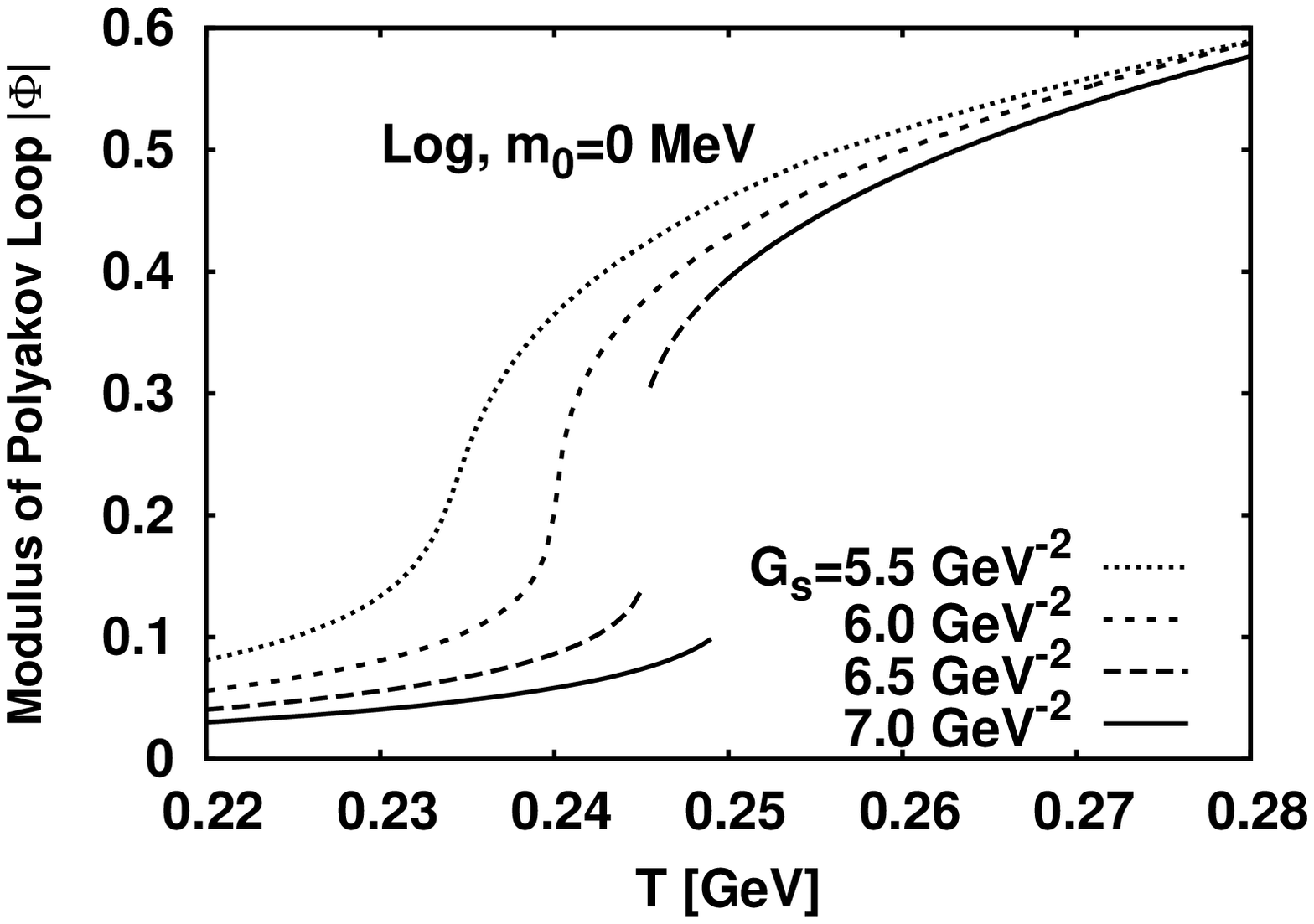,width=0.45\textwidth}
 \epsfig{file=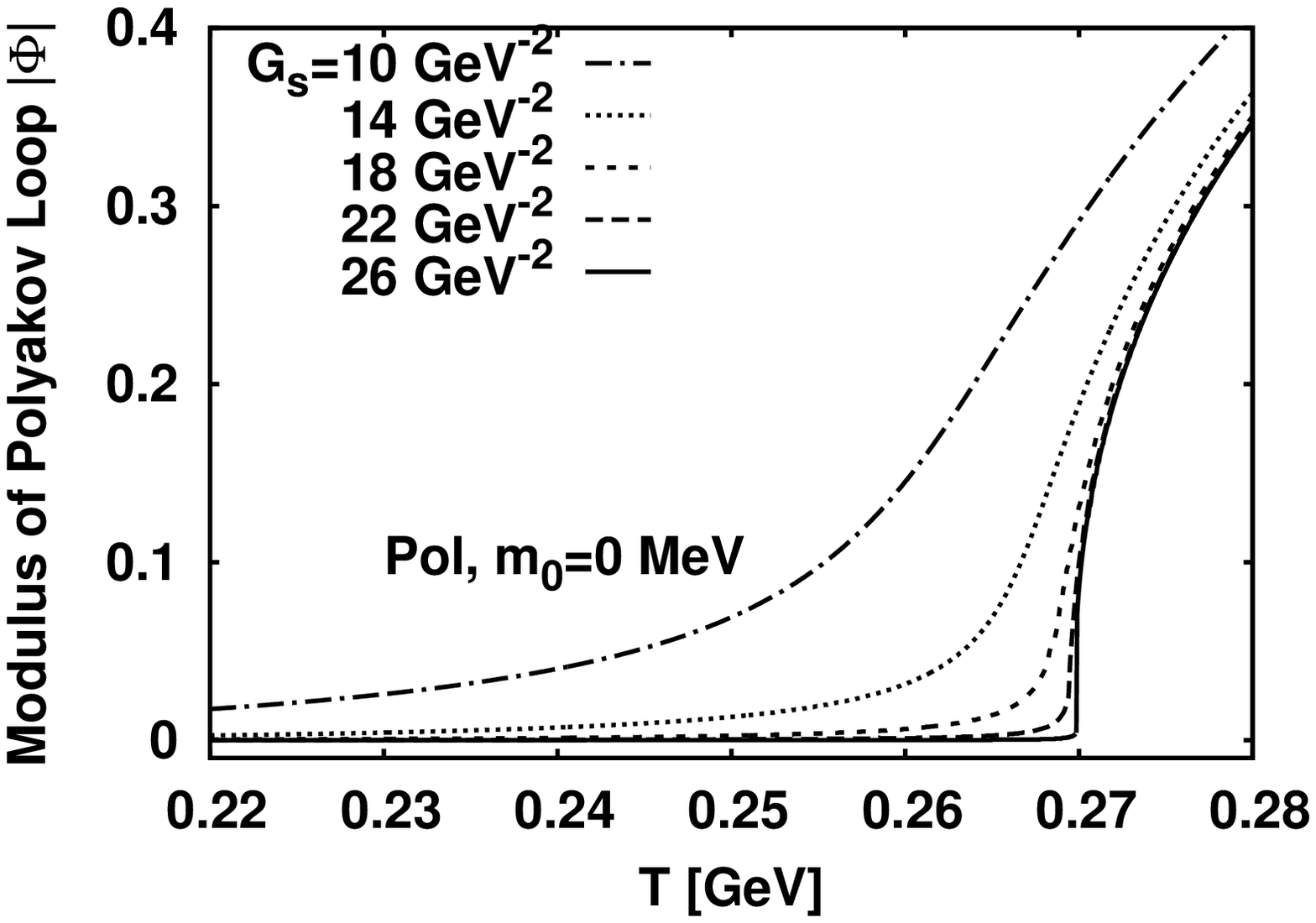,width=0.45\textwidth}
 \caption{Behavior of $|\Phi|$ for large $G_s$. Left : 
 $\mathcal{U}_{\rm  log}$. Right : $\mathcal{U}_{\rm poly}$}
 \label{fig:polgs}
\end{figure}

Figure \ref{fig:polgs} shows the behavior of $|\Phi|$ for various $G_s$ at vanishing
chemical potential. One sees that $|\Phi|$ becomes steeper for larger
$G_s$ in both of $\mathcal{U}$.
The case of $\mathcal{U}_{\rm log}$  has a discontinuity already at
$G_s=6.5$ GeV$^{-2}$ as mentioned above. $\mathcal{U}_{\rm pol}$, which has a smoother variation of $|\Phi|$
against $T$, eventually approaches the pure gauge case for much
larger $G_s$. At $G_s=25$ GeV$^{-2}$,
where the dynamical quark mass at $T=0$ is around 2.5 GeV, the first
order deconfinement transition is recovered. The origin of this difference
is the much weaker first order transition in $\mathcal{U}_{\rm poly}$,
which easily turns into crossover when quarks heavier than 2.5 GeV are
put into the system. If one characterizes a strength of the
deconfinement transition by a gap of the Polyakov loop $\Delta \Phi$ at $T=T_0$, 
one finds $\Delta \Phi=0.47$ for $\mathcal{U}_{\rm log}$ and 0.072 for
$\mathcal{U}_{\rm poly}$. Since $G_s$ determines the scale of the
dynamical chiral symmetry breaking, $\sigma(T=0)$, this result indicates
an interplay of the two transitions which have a unique scale,
$\Lambda_{\rm QCD}$,  in the case of QCD.

It has been shown that a first order deconfinement phase transition also emerges in the large
$N_c$ limit of the PNJL model
\cite{mclerran09:_quark_matter_and_chiral_symmet_break}. This result
has a common origin with the present study in a sense that taking large
$N_c$ limit makes the system gluon dominated due to $1/N_c$ suppression of the quark
contribution, while large $G_s$  thermally suppresses quarks in
the chirally broken phase. In our case, however, chiral transition
temperature moves upward thus there is a discrepancy between the
deconfinement and chiral transition temperatures, in contrast to the
large $N_c$ limit with a fixed $G_s N_c$ in \cite{mclerran09:_quark_matter_and_chiral_symmet_break}.

\section{Summary}
\label{sec:summary}

We have explored the deconfinement transition in the PNJL model at imaginary
chemical potential. We point out that the chiral condensate at imaginary
chemical potential, $\sigma(\theta)$, has a characteristic $\theta$
dependence due to the deconfinement property which naturally arises from the
statistical confinement feature of the model. While the confined
phase is characterized by a smooth $\cos3\theta$ dependence, the
deconfined phase exhibits $\cos\theta$ dependence together with cusps at
$\theta=\pi/3$ (mod $2\pi/3$) induced by the abrupt change of the phase of
the Polyakov loop (Roberge-Weiss transition). 
We introduce a new dual parameter utilizing this $\theta$ dependence and
compare it with the Polyakov loop and the dressed Polyakov loop. 
Different sensitivities of these parameters to chiral and
deconfinement transitions are found. 
Changing the four fermion coupling constant, we found that an
interplay between the thermal quark contribution through the dynamical
chiral symmetry breaking and the Polyakov loop potential
determines the location of the deconfinement CEP at imaginary chemical
potential. In particular, we found that the deconfinement CEP can be located
in the real chemical potential regime for a Polyakov loop potential with a strong
first order transition and large dynamical chiral symmetry breaking.
We expect that these results are useful for understanding of the QCD
phase transition.

This work is supported by the Yukawa International Program for
Quark-Hadron Sciences at Kyoto University. K.M. and V.S. acknowledges
FIAS for support. B.F. and K.R. acknowledges partial support by
EMMI. K.R. acknowledges partial support by the Polish Ministry of
Science (MEN).V.S. was supported by the U.S. Department of Energy under
Contract No. DE-AC02-98CH10886.

%\bibliography{adsqcd,charm,chiral,chpt,eos,exotic,experiment,hydro,jpsi_sup,lattice,morita,nuclearmatter,pdg,qcd,qgp,QGPreview,qm08,sumrule,textbook,tft,thermalmodel,PNJL,PQM}

\begin{thebibliography}{10}
%\expandafter\ifx\csname url\endcsname\relax
%  \def\url#1{\texttt{#1}}\fi
%\expandafter\ifx\csname urlprefix\endcsname\relax\def\urlprefix{URL }\fi
%\expandafter\ifx\csname href\endcsname\relax
%  \def\href#1#2{#2} \def\path#1{#1}\fi

\bibitem{aoki06}
	Y.~Aoki, G.~Endr\"{o}di, Z.~Fodor, S.~D. Katz, K.~K. Szab\'{o}, Nature
	{\bf 443}, 675 (2006); A.~Bazavov {\it et al.}, arXiv:1111.1710.

\bibitem{muroya03}
	S.~Muroya, A.~Nakamura, C.~Nonaka, and T.~Takaishi,
	Prog. Theor. Phys. {\bf 110}, 615 (2003).

\bibitem{forcrand02:_qcd}
	P.~de~Forcrand, O.~Philipsen, Nucl. Phys. {\bf B642}, 290 (2002); {\it
	ibid}, {\bf B673} 170 (2003); J. High. Energy Phys. {\bf 01} 077
	(2007).

\bibitem{d'elia-lombardo}
	M.~D'Elia and M.~P.~Lombardo, Phys. Rev. {\bf D67}, 014505
	(2003); {\it ibid}, {\bf D70}, 074509 (2004).

\bibitem{d'elia07}
	M.~D'Elia, F.~D.~Renzo, and M.~P.~Lombardo, Phys. Rev. {\bf 76},
	114509 (2007).

\bibitem{roberge86:_gauge_qcd}
A.~Roberge, N.~Weiss, Nucl. Phys. {\bf B275}, 734 (1986).

\bibitem{forcrand:_const_qcd}
P.~de~Forcrand, O.~Philipsen, Phys. Rev. Lett. {\bf 105}, 152001 (2010).
%\newblock \href {http://arxiv.org/abs/arXiv:1004.3144}
%  {\path{arXiv:arXiv:1004.3144}}.

\bibitem{bonati:_rober_weiss_endpoin_in_n_f_qcd}
M.~D'Elia and F.~Sanfilippo, Phys. Rev. {\bf D80}, 111501(R) (2009);
C.~Bonati, G.~Cossu, M.~D'Elia, F.~Sanfilippo, Phys. Rev. {\bf D83}, 054505 (2011).
%\newblock \href {http://arxiv.org/abs/arXiv:1011.4515}
%  {\path{arXiv:arXiv:1011.4515}}.

\bibitem{Chen}
	H.~S.~Chen and X.~Q.~Luo, Phys. Rev. {\bf D72}, 034504 (2005);
	L.~K.~Wu, X.~Q.~Luo, and H.~S.~Chen, Phys. Rev. {\bf D76},
	034505 (2007).

\bibitem{Nagata}
	K.~Nagata and A.~Nakamura, Phys. Rev. {\bf D83}, 114507 (2011).

\bibitem{fukushima04:_chiral_polyak}
K.~Fukushima, Phys. Lett. {\bf B591}, 277 (2004).

\bibitem{Boyd}
	G.~Boyd et al., Nucl. Phys. {\bf B469}, 419 (1996).
%, J.~Engles, F.~Karsch, E.~Laermann, C.~Legeland,
%	M.~L\"{u}tgemeier, and B.~Petersson, Nucl. Phys. {\bf B469}, 419
%	(1996).

\bibitem{ratti06:_phases_qcd}
C.~Ratti, M.~A. Thaler, W.~Weise, Phys. Rev. {\bf D73}, 014019 (2006).

\bibitem{morita11:_probin_decon_in_chiral_effec}
K.~Morita, V.~Skokov, B.~Friman, K.~Redlich, Phys. Rev. {\bf D84}, 076009 (2011).
%\newblock \href {http://arxiv.org/abs/1107.2273} {\path{arXiv:1107.2273}}.

\bibitem{nambu61:_NJLI}
Y.~Nambu, G.~Jona-Lasinio, Phys. Rev. {\bf 122}, 345 (1961); {\it
	ibid}. {\bf 124}, 246 (1961).

\bibitem{hatsuda94:_qcd_lagran}
T.~Hatsuda, T.~Kunihiro, Phys. Rept. {\bf 247}, 221 (1994).

\bibitem{roessner07:_polyak}
S.~Roessner, C.~Ratti, W.~Weise, Phys. Rev. {\bf D75}, 034007 (2007).

\bibitem{sakai08:_polyak_nambu_jona_lasin}
Y.~Sakai, K.~Kashiwa, H.~Kouno, M.~Yahiro, Phys. Rev. {\bf D77}, 051901(R) (2008).

\bibitem{sakai:_10063408}
Y.~Sakai, T.~Sasaki, H.~Kouno, M.~Yahiro, Phys. Rev. {\bf D82}, 076003 (2010).
%\newblock \href {http://arxiv.org/abs/arXiv:1006.3408}
%  {\path{arXiv:arXiv:1006.3408}}.

\bibitem{bilgici08:_dual_polyak}
E.~Bilgici, F.~Bruckmann, C.~Gattringer, C.~Hagen, Phys. Rev. {\bf D77},
	094007 (2008).

\bibitem{weiss87:_how}
N.~Weiss, Phys. Rev. {\bf D35}, 2495 (1987).

\bibitem{kashiwa09:_dual_polyak_nambu_jona_lasin}
K.~Kashiwa, H.~Kouno, M.~Yahiro, Phys. Rev. {\bf D80}, 117901 (2009).

\bibitem{mclerran09:_quark_matter_and_chiral_symmet_break}
L.~McLerran, K.~Redlich, C.~Sasaki, Nucl. Phys. {\bf A824}, 86 (2009).

\end{thebibliography}
\end{document}